\documentclass[aps,prl,twocolumn,amsmath,amssymb,floatfix]{revtex4}
\usepackage{graphicx}
\begin{document}

% Use the \preprint command to place your local institutional report
% number in the upper righthand corner of the title page in preprint mode.
% Multiple \preprint commands are allowed.
%\preprint{}

\title{Exact and Approximate Performance of Concatenated Quantum Codes}

\author{Benjamin Rahn} 
\email{brahn@caltech.edu} 
\author{Andrew C. Doherty}
\author{Hideo Mabuchi}
\affiliation{Institute for Quantum
Information, California Institute of Technology}
\date{November 1, 2001}

\begin{abstract}
We derive the effective channel for a logical qubit protected by an
arbitrary quantum error-correcting code, and derive the map between
channels induced by concatenation.  For certain codes in the presence of
single-bit Pauli errors, we calculate the exact threshold error
probability for perfect fidelity in the infinite concatenation limit.
We then use the control theory technique of balanced truncation
to find low-order non-asymptotic approximations for the effective channel
dynamics. 
\end{abstract}

% insert suggested PACS numbers in braces on next line

%\maketitle must follow title, authors, abstract, \pacs, and \keywords
\maketitle

\def\expval[#1]{\langle #1 \rangle}
\def\tr{\mathrm{tr}}
\def\comment{\bf \footnotesize}
\def\ket[#1]{\vert #1 \rangle}
\def\bra[#1]{\langle #1 \vert}
\def\braket[#1,#2]{\langle #1 \vert #2 \rangle}
\def\ketbra[#1,#2]{\vert #1 \rangle \langle #2 \vert}
\def\Eop{\mathcal{E}}
\def\Gop{\mathcal{G}}
\def\Gdep{\mathcal{G}^\mathrm{dep}}
\setlength{\parskip}{0pt}
\setlength{\textfloatsep}{5pt}
\setlength{\floatsep}{6pt}
\def\fbar{{\bar{f}}}
\def\hide[#1]{}

Quantum Error Correction \cite{qecc_standard} and Fault-Tolerant
Computation \cite{ftc_standard} have demonstrated that, in principle,
quantum computing is possible despite noise in the computing device.
Analyses in these areas, as in many physics disciplines, rely on
asymptotic limits and expansions in small parameters.  However,
realistic devices will be of large but finite scale and real parameter
values may not be sufficiently small; thus methods valid outside these
limits are desirable.  Here we demonstrate the use of \textit{model
reduction} \cite{mr_standard,mr_primer}, a control theory technique, for
studying large but finite quantum systems.

We will first derive the effective channel describing an encoded qubit's
evolution under arbitrary error dynamics, and propagate
these results to concatenation schemes \cite{qecc_standard} when the
dynamics do not couple code blocks.  The calculations for both finite
and asymptotic concatenation are simple, but the resulting expressions for the
effective channel dynamics in the finite case are cumbersome and
ill-conditioned.  We will then use model reduction to find low-order
approximations for these dynamics.  Among other results, we derive a
model for a $9^4$-qubit system requiring only 23 degrees of freedom for
accuracy $\sim 10^{-3}$.

We consider the following formulation of quantum error correction: 
a single-qubit state $\rho_0$ is perfectly encoded in a multi-qubit
register with state $\rho$, which evolves via some error dynamics
described by a linear map $\rho(0) \mapsto \rho(t) = \Eop[\rho(0)]$.
(For master equation evolution $\dot{\rho} = \mathcal{L}[\rho]$, $\Eop =
e^{\mathcal{L}t}$.)  At time $t$, a syndrome measurement is made and the
appropriate recovery operator is performed; we assume the measurement
and recovery are noiseless.  This process returns the system to the
codespace \footnote{We may choose such a recovery procedure even if the
code is imperfect.}, and thus its state can be described by a
single-qubit state $\rho_f$. Denote the probability-weighted average of
$\rho_f$ over syndrome measurement outcomes by $\overline{\rho_f}$.  For
a given code and error model, we wish to know $\overline{\rho_f}(t)$: we
may then compare $\rho_0$ and $\overline{\rho_f}(t)$ via a desired
fidelity measure.

The logical qubit $\rho_0$ may be parameterized by the expectation
values $\expval[I]_0$, $\expval[X]_0$, $\expval[Y]_0$ and
$\expval[Z]_0$, with $\{I,X,Y,Z\}$ the usual Pauli matrices.  (Of course
$\expval[I]=1$, but it will be convenient to include this term.)  The
logical qubit $\rho_0$ is encoded by preparing the register in the state $
\rho(0) = \expval[I]_0 E_I + \expval[X]_0 E_X + \expval[Y]_0 E_Y +
\expval[Z]_0 E_Z $ where $E_\sigma$ acts as $\frac{1}{2}\sigma$ on the
two-dimensional codespace and vanishes elsewhere.  These operators are
easily constructed from the codewords: for example, given the encoding
$\ket[0]\mapsto\ket[\overline{0}]$, $\ket[1]\mapsto\ket[\overline{1}]$,
we have $E_X = \frac{1}{2}(\ketbra[\overline{0},\overline{1}] +
\ketbra[\overline{1},\overline{0}])$. (For a stabilizer code
\cite{qecc_standard} with stabilizer $S = \{S_i\}$ and logical
operators $\bar{\sigma}$, the codespace projector is $P_C =
\frac{1}{|S|}\sum_i S_i$ and $E_\sigma = \frac{1}{2} P_C \bar{\sigma}$.)
E.g., for the bitflip code \cite{qecc_standard} given by
$\ket[0]\mapsto\ket[000]$, $\ket[1]\mapsto\ket[111]$,
{\setlength{\arraycolsep}{2pt}
\begin{equation}
\label{eqn:bit_enc_ops}
\begin{array}{rclcccccccl}
E_I &=&\frac{1}{8}( & III  & + & IZZ & + & ZIZ & + & ZZI & ) \\ 
E_X &=&\frac{1}{8}( & XXX  & - & XYY & - & YXY & - & YYX & ) \\
E_Y &=&\frac{1}{8}( & -YYY & + & YXX & + & XYX & + & XXY & ) \\
E_Z &=&\frac{1}{8}( & ZZZ  & + & ZII & + & IZI & + & IIZ & ).
\end{array}
\end{equation}}
For the trivial code ($\rho_0$ ``encoded'' as itself in a single-qubit
register) we have simply $E_\sigma = \frac{1}{2}\sigma$.  

After the action of $\Eop$, recovery yields the expected logical state
$\overline{\rho_f}$.  As with $\rho_0$, we parameterize
$\overline{\rho_f}$ by the expectation values $\expval[I]_{\fbar}$,
$\expval[X]_{\fbar}$, $\expval[Y]_{\fbar}$ and $\expval[Z]_{\fbar}$.
The $\expval[\sigma]_{\fbar}$ may be written as expectation values of
operators on the register \textit{prior} to recovery: letting $\{P_j\}$
be the syndrome measurement projectors and $\{R_j\}$ be the recovery operators,
$\expval[\sigma]_{\fbar} = \mathrm{tr}(D_\sigma\rho)$ where $D_\sigma =
2\sum_j P_j^\dagger R_j^\dagger E_\sigma R_j P_j$. (For a stabilizer
code, $R_j$ is some Pauli operator of lowest weight leading to syndrome
measurement $P_j$, and $D_\sigma = \frac{1}{|S|}\sum_i f_{i\sigma}S_i
\bar{\sigma}$, with $f_{i\sigma} =
\sum_j\eta(S_i,R_j)\eta(R_j,\bar{\sigma})$ where $\eta(p,q) = \pm 1$ for
$pq = \pm qp$.)  For the bitflip code, {\setlength{\arraycolsep}{2pt}
\begin{equation}
\label{eqn:bit_dec_ops}
\begin{array}{ccrcccccccl}
D_I &=& & III \\
D_X &=& & XXX \\
D_Y &=& \frac{1}{2}( & YYY & + & YXX & + & XYX & + & XXY &) \\
D_Z &=& \frac{1}{2}( & -ZZZ & + & ZII & + & IZI & + & IIZ &).
\end{array}
\end{equation}}
For the trivial code we have simply $D_\sigma = \sigma$.  

To describe the evolution of the encoded logical bit, we compute the
effective channel $\Gop$ taking $\rho_0$ to $\overline{\rho_f}$.  $\Gop$
may be written as the linear mapping of $\vec{\rho}_0 =
(\expval[I]_0,\expval[X]_0,\expval[Y]_0,\expval[Z]_0)$ to
$\vec{\rho}_\fbar =
(\expval[I]_\fbar,\expval[X]_\fbar,\expval[Y]_\fbar,\expval[Z]_\fbar)$:
\begin{equation}
\expval[\sigma]_\fbar = \tr \left ( D_\sigma\Eop \left [ \sum_{\sigma'}
                        \expval[\sigma']_0 E_{\sigma'} \right ] \right
                        ).
\end{equation}
If $\Eop$ is completely positive \cite{qecc_standard}, 
it follows that $\Gop$ is as well.

To obtain a matrix representation of $\Gop$, let
\begin{equation}
\label{eqn:G_formula}
\Gop_{\sigma \sigma'} = \tr ( D_\sigma \Eop 
                           \left [ E_{\sigma'} \right ]).
\end{equation}
Then $\vec{\rho}_\fbar = \Gop \vec{\rho}_0$, and the fidelity of a pure
logical qubit under this process is $\frac{1}{2}\vec{\rho}_0^{\:{\mathrm T}}
\Gop \vec{\rho}_0$.  Note that the dynamics $\Eop$ need not be those
against which the code protects.

Now consider concatenated codes \cite{qecc_standard}.  In the
concatenation of two codes, a qubit is encoded using the outer code
$C^\mathrm{out}$ and then each of the resulting qubits is encoded using
the inner code $C^\mathrm{in}$.  Though not necessarily optimal, a simple
error-correction scheme coherently corrects each of the inner code
blocks, and then corrects the entire register based on the outer code.
We denote the concatenated code (with this correction scheme) by
$C^\mathrm{out}(C^\mathrm{in})$.  Given the effective channel $\Gop$ 
describing $C^\mathrm{in}$ under some dynamics, and a desired
$C^\mathrm{out}$, we now construct $\widetilde{\Gop}$ describing the
evolution under $C^\mathrm{out}(C^\mathrm{in})$.

Let $C^\mathrm{out}$ be an $N$-bit code and $C^\mathrm{in}$ be an $M$-bit
code.  We assume that each $M$-bit block evolves according to the
original dynamics $\Eop$ and no cross-block correlations are
introduced, thus the evolution operator is $\widetilde{\Eop} = \Eop
\otimes \Eop \otimes \ldots \otimes \Eop$.  Each $M$-bit block
represents a single logical qubit encoded in $C^\mathrm{in}$; as the block
has dynamics $\Eop$, this logical qubit's evolution is described by
$\Gop$.  Therefore the effective evolution operator for the logical bits
in the codeword of $C^\mathrm{out}$ is $\widetilde{\Eop}_\mathrm{eff} = \Gop
\otimes \Gop \otimes \ldots \otimes \Gop.$

Operators on $N$ qubits may be written as sums of tensor products of
$N$ Pauli matrices; we may therefore write
\begin{eqnarray}
\label{eqn:E_D_out}
E_{\sigma'}^{\mathrm{out}} &=& 
\sum_{\genfrac{}{}{0pt}{}{\mu_i \in}{\{I,X,Y,Z\}}}
\alpha^{\sigma'}_{\{\mu_i\}}
\left (\tfrac{1}{2} \mu_1 \right) 
\otimes \ldots \otimes 
\left (\tfrac{1}{2} \mu_N \right) 
\\
D_\sigma^{\mathrm{out}} &=& \sum_{ 
\genfrac{}{}{0pt}{}{\nu_i \in}{\{I,X,Y,Z\}}}
\beta^\sigma_{\{\nu_i\}}
\nu_1 \otimes \ldots \otimes \nu_N.
\end{eqnarray}
(For stabilizer codes, the $\alpha$ and
$\beta$ coefficients are easily found.)  Substituting
$\widetilde{\Eop}_\mathrm{eff}$, $E^\mathrm{out}_{\sigma'}$ and
$D^\mathrm{out}_\sigma$ into (\ref{eqn:G_formula}) and noting that
$\tr(\nu_j \Gop[\frac{1}{2}\mu_j]) = \Gop_{\nu_j \mu_j}$ yields
\begin{equation}
\label{eqn:G_concat}
\widetilde{\Gop}_{\sigma \sigma'} =
\sum_{\{\mu_i\},\{\nu_i\}}
\left (
\beta_{\{\nu_i\}}^{\sigma}
\alpha_{\{\mu_i\}}^{\sigma'}
\prod_{j = 1}^N \Gop_{{\nu_j}{\mu_j}} \right ).
\end{equation}
Thus the matrix elements of $\widetilde{\Gop}$ can be
expressed as polynomials of the matrix elements of $\Gop$, with the
polynomial coefficients $\beta_{\{\nu_i\}}^{\sigma}
\alpha_{\{\mu_i\}}^{\sigma'}$ depending only on the $E_\sigma$ and
$D_\sigma$ of the outer code.  For a given outer code $C$, denote the
concatenation map $\Gop \mapsto \widetilde{\Gop}$ by $\Omega^C$.
It can be shown that $\Omega^C$ preserves complete positivity.

More generally, the inner process $\Gop$ can represent any linear
evolution of a logical qubit, and thus we may speak of concatenating a
qubit process with a code.  E.g., if $\Gop$ describes some qubit
dynamics, then $\Omega^C(\Gop)$ describes the effective dynamics of
encoding by $C$ with the uncorrelated dynamics $\Gop$ acting on each
register bit.  This method only requires that the outer code's logical
qubits be decoupled.  (Above we assumed $\widetilde{\Eop}_\mathrm{eff} =
\Gop \otimes \ldots \otimes \Gop$; for $\widetilde{\Eop}_\mathrm{eff}
= \Gop^{(1)} \otimes \ldots \otimes \Gop^{(N)}$, replace $\Gop_{\nu_j
\mu_j}$ with $\Gop^{(j)}_{\nu_j \mu_j}$ in (\ref{eqn:G_concat}).)

We may characterize both the finite and asymptotic behavior of any
concatenation scheme involving the codes $\{C_k\}$ by computing the maps
$\Omega^{C_k}$.  Then the finite concatenation scheme $C_1(C_2(\ldots
C_n\ldots))$ is characterized by $\Omega^{C_1(C_2(\ldots C_n\ldots))} =
\Omega^{C_1}(\Omega^{C_2}(\ldots \Omega^{C_n}\ldots))$.  We
expect the typical $\Omega^C$ to be sufficiently well-behaved that standard
dynamical systems methods \cite{dynamical_standard} will yield the $\ell
\rightarrow \infty$ limit of $(\Omega^C)^\ell$; one need not compose the
$(\Omega^C)^\ell$ explicitly.

We now consider certain concatenation schemes when the symmetric
depolarizing channel \cite{qecc_standard} acts on each register qubit.
This channel is described by $\Gdep(t)$ diagonal with entries
$(1,e^{-\gamma t},e^{-\gamma t},e^{-\gamma t})$.  From trace
preservation $\Gop_{II}$ is always 1, so let $[x,y,z]$ denote
$\Gop$ diagonal with entries $(1,x,y,z)$; then $\Gdep(t) =
[e^{-\gamma t},e^{-\gamma t},e^{-\gamma t}]$.

Suppose more generally we are given a qubit process described by $\Gop =
[x,y,z]$, and wish to concatenate this process with the bitflip (bf) code.
Using (\ref{eqn:G_concat}) and the coding operators
(\ref{eqn:bit_enc_ops}) and (\ref{eqn:bit_dec_ops}), we find that
$\Omega^\mathrm{bf}(\Gop)$ describing the concatenated evolution is also
diagonal:
\begin{equation}
\label{eqn:poly_bit}
\Omega^\mathrm{bf}([x,y,z])=
\left [ x^3,\: \tfrac{3}{2}x^2 y-\tfrac{1}{2}y^3, 
            \:  \tfrac{3}{2}z-\tfrac{1}{2}z^3 \right ].
\end{equation}
((\ref{eqn:poly_bit}) could also be found by using the Heisenberg
picture to evaluate (\ref{eqn:G_formula}).)  Writing $\ket[\pm] =
\frac{1}{\sqrt{2}}(\ket[0] \pm \ket[1])$, the map $\Omega^\mathrm{pf}$ for
the phaseflip code $\ket[\pm] \mapsto \ket[\!\pm\!\pm\pm]$
\cite{qecc_standard} is similar.  (Any stabilizer code $C_S$ preserves
diagonality: 
$\Omega^{C_S}([x,y,z])$ is
\begin{equation}
\widetilde{\Gop}_{\sigma\sigma'}=
\delta_{\sigma\sigma'}
\frac{1}{|S|}\sum_i f_{i\sigma}
x^{w_X(S_i\bar{X})} y^{w_Y(S_i\bar{Y})}
z^{w_Z(S_i\bar{Z})}
\end{equation}
with $w_\sigma(p)$ the $\sigma$-weight of a Pauli
operator $p$ (e.g. $w_X(XYX)=2$)
and $f_{i\sigma}$ as previously defined.)

The concatenation phaseflip(bitflip) yields the encoding $\ket[\pm]
\mapsto \frac{1}{\sqrt{8}}(\ket[000] \pm \ket[111])^{\otimes 3}$,
which is the Shor nine-bit code \cite{qecc_standard}.  Thus
$\Omega^\mathrm{Shor}=\Omega^\mathrm{pf}(\Omega^\mathrm{bf})$, and
$\Omega^\mathrm{Shor}([x,y,z]) =$
\begin{eqnarray}
\label{eqn:poly_shor}
\left [ \tfrac{3}{2}x^3 - \tfrac{1}{2}x^9, \: 
\tfrac{3}{2}\left (\tfrac{3}{2}z - \tfrac{1}{2}z^3 \right )^2
  \left (\tfrac{3}{2}x^2 y - \tfrac{1}{2}y^3 \right ) \right.
\nonumber \\
\left. -\tfrac{1}{2}\left (\tfrac{3}{2}x^2 y - \tfrac{1}{2}y^3 \right )^3, \:
\left (\tfrac{3}{2}z - \tfrac{1}{2}z^3 \right )^3 \right ] .
\end{eqnarray}

\begin{figure}
\includegraphics[scale=0.47]{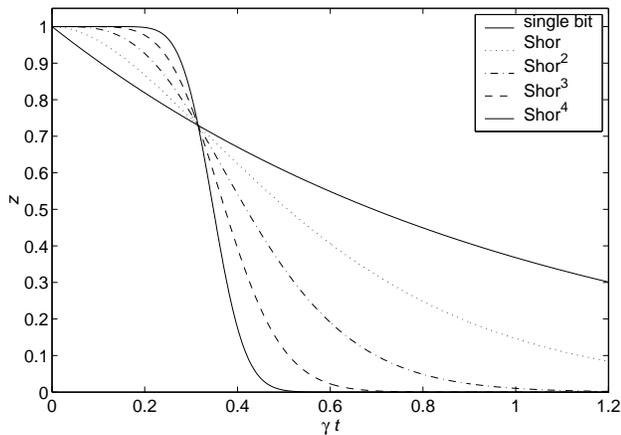}
\caption{$\widetilde{z}_\ell(t)$ for Shor$^\ell$ concatenation under the
depolarizing channel; the fidelity of an encoded $Z$ eigenstate is
$\frac{1}{2}(1+\widetilde{z}_\ell(t))$.
\label{fig:shor_pointwise_Z}}
\end{figure}

Now consider the Shor code concatenated with itself $\ell$ times, and
let $[\widetilde{x}_\ell(t),\widetilde{y}_\ell(t),\widetilde{z}_\ell(t)]
=(\Omega^\mathrm{Shor})^\ell(\Gdep(t))$.  The functions
$\widetilde{\sigma}_\ell(t)$ approach step functions in the limit $\ell
\rightarrow \infty$ (e.g., see Fig. \ref{fig:shor_pointwise_Z}); denote
these step functions' times of discontinuity by $t_\sigma^\star$.  Thus in the
infinite concatenation limit, the code will perfectly protect the
$\expval[\sigma]$ component of the logical qubit if correction is
performed prior to $t_\sigma^\star$.  We call $t_\sigma^\star$ the
$\sigma$-\textit{storage threshold}.

We calculate the $t_\sigma^\star$ by finding the $\ell \rightarrow
\infty$ limit of $\widetilde{\sigma}_\ell(t)$.  Writing
(\ref{eqn:poly_shor}) as $[Q_1(x)$, $Q_2(x,y,z),Q_3(z)]$, the map $z
\mapsto Q_3(z)$ has stable fixed points at 0 and 1 and one unstable
fixed point $z^\star$ on $(0,1)$; numerically solving $Q_3(z) = z$
yields $z^\star \approx 0.730$.  The plots of $\widetilde{z}_\ell(t)$
all intersect at $\widetilde{z}(t_Z^\star) = z^\star$, and the step
function limit follows from the stability of 0 and 1.  Inverting
$e^{-\gamma t} = z^\star$ yields $t_Z^\star$.  $Q_1$ has similar
features with $x^\star \approx 0.900$, and a similar analysis yields
$t_X^\star$.  For this code, one can show $t_Y^\star =
\min(t_X^\star,t_Z^\star)$.

We may also phrase the thresholds in the language of finitely probable
errors.  The expected evolution of a qubit subjected to a random Pauli
error with probability $p$ is $\rho \mapsto (1-p)\rho +
\frac{p}{3}X\rho X + \frac{p}{3}Y\rho Y + \frac{p}{3}Z\rho Z$.  This
channel is described by $\Gop^\mathrm{Pauli}(p) = [1 - \frac{4}{3}p, 1 -
\frac{4}{3}p, 1 - \frac{4}{3}p]$.  As
$\Gop^\mathrm{Pauli}(\frac{3}{4}(1-e^{-\gamma t})) = \Gdep(t)$, in the
infinite concatenation limit with $\Gop^\mathrm{Pauli}(p)$ acting on each
register qubit, the logical qubit's $\expval[\sigma]$ component will be
perfectly protected if $p < p^\star_\sigma = \frac{3}{4}(1-e^{-\gamma
t^\star_\sigma})$.  Define the threshold probability $p_\mathrm{th} =
\min\{p^\star_\sigma\}$; for $p < p_\mathrm{th}$, all encoded qubits are
perfectly protected in the infinite concatenation limit.  Values for
$t^\star_\sigma$ and $p_\mathrm{th}$ appear in Table
\ref{tbl:asymptotics}.

For comparison, we derived thresholds for three other codes.  Another
version of the Shor code is given by $\ket[0] \mapsto
\frac{1}{\sqrt{8}}(\ket[000]+\ket[111])^{\otimes 3}$, $\ket[1]
\mapsto \frac{1}{\sqrt{8}}(\ket[000]-\ket[111])^{\otimes 3}$; call
this code Shor$'$.  Let
$[\widetilde{x}'_\ell(t),\widetilde{y}'_\ell(t),\widetilde{z}'_\ell(t)]
=(\Omega^{\mathrm{Shor}'})^\ell(\Gdep(t))$.  The $\widetilde{y}'_\ell(t)$
approach a step function as $\ell
\rightarrow \infty$, but $\widetilde{x}'_\ell(t)$ and
$\widetilde{z}'_\ell(t)$ approach a limit cycle of period 2,
interchanging step functions with different discontinuities at every
iteration of $\Omega^{\mathrm{Shor}'}$.  Considering instead the limit of
iterating $(\Omega^{\mathrm{Shor}'})^2$ permits an analysis as for
$\Omega^\mathrm{Shor}$.  The Steane seven-bit code \cite{qecc_standard}
may be treated similarly to the Shor code, and the symmetries of the
Five-Bit code \cite{qecc_standard} lead to a simple analysis.  Results
are summarized in Table \ref{tbl:asymptotics}.

\begin{table}
\begin{tabular}{|c||c|c||c|c||c||c|}
\hline
Code & \multicolumn{2}{|c||}{Shor} & \multicolumn{2}{|c||}{Shor$'$} 
     & Steane & Five-Bit\\
\hline
$\sigma$ & $X, Y$ & $Z$ & $X, Y$ & $Z$ & $X,Y,Z$ & $X,Y,Z$\\
\hline
$\gamma t_\sigma^\star$ & 0.1050 & 0.3151 & 0.1618 &  0.2150 & 0.1383 & 0.2027 \\
\hline
$p_\mathrm{th}$ & \multicolumn{2}{|c||}{0.0748} &
\multicolumn{2}{|c||}{0.1121} & 0.0969 & 0.1376
\\
\hline
\end{tabular}
\caption{Code storage thresholds.
\label{tbl:asymptotics}}
\end{table}

We now return to the Shor code under the depolarizing channel, and
consider the finite concatenations described by the functions
$\widetilde{\sigma}_\ell(t)$.  The $\widetilde{\sigma}_\ell(t)$ have the
form $\sum_i b_i e^{-a_i \gamma t}$ with the $a_i$ positive integers and
the $b_i$ rationals.  For $\gamma = 0$ no errors occur, thus $\sum_i b_i
= 1$.

Explicit calculation of the $\widetilde{\sigma}_\ell$ has several
disadvantages.  First, the number of terms in these series grows
approximately as $9^\ell$ (see Table \ref{tbl:terms_and_order}(a)).
Though not nearly as severe as for the number of elements in the
full-system density matrix ($2^{2 \cdot 9^\ell}$), this growth is still
too rapid to be practical.  Only a small portion of the terms in these
series have $|b_i| < 1$, thus one cannot meaningfully truncate the
series without introducing significant error.

\begin{table}
\begin{tabular}{|c||c|c|c||c|c|c|}
\hline
$\ell$ (qubits) & \multicolumn{3}{|c||}{series terms} &
                  \multicolumn{3}{|c|}{reduced order} \\
\hline
& $\widetilde{x}_\ell$  & $\widetilde{y}_\ell$ & 
                               $\widetilde{z}_\ell$ & 
{\hbox to18pt{\hfil$\widetilde{x}_\ell$\hfil}} & 
{\hbox to18pt{\hfil$\widetilde{y}_\ell$\hfil}} & 
{\hbox to18pt{\hfil$\widetilde{z}_\ell$\hfil}} \\
\hline
0 (1)  & 1 & 1 & 1 & 1 & 1 & 1 \\
\hline
1 (9)  & 2 & 3 & 4 & 2 & 2 & 3 \\
\hline
2 (81) & 13 & 33 & 37 & 4 & 4 & 5 \\
\hline
3 (729) & 118 & 339 & 352 & 5 & 5 & 6 \\
\hline
4 (6561) & 1081 & 3201 & 3241 & 7 & 7 & 9\\
\hline
&\multicolumn{3}{|c||}{(a)}  & \multicolumn{3}{|c|}{(b)}  \\
\hline
\end{tabular}
\caption{(a) Terms in exact series for $\widetilde{\sigma}_\ell(t)$. (b)
Order of iteratively reduced realizations for
$\widetilde{\sigma}_\ell(t)$.
\label{tbl:terms_and_order}}
\end{table}

More seriously, the magnitude of the $b_i$ grows rapidly: e.g., $|b_i| >
10^{60}$ for 65 of the 352 terms in $\widetilde{z}_3$, and
double-floating point precision no longer yields $\sum_i b_i = 1$. To
efficiently generate plots of the $\widetilde{\sigma}_\ell(t)$ we
repeatedly apply $\Omega^\mathrm{Shor}$ to \textit{numerical} values of
$[e^{-\gamma t}, e^{-\gamma t},e^{-\gamma t}]$ for all desired times
$t$.  However, this leaves us without a dynamic model for the
evolution of $\overline{\rho_f}(t)$.

Given $\Gop(t) = [\widetilde{x}(t),\widetilde{y}(t),\widetilde{z}(t)]$,
for each $\widetilde{\sigma}(t)$ we will seek a square matrix
$A_\sigma$, column vector $B_\sigma$ and row vector $C_\sigma$ such that
$\widetilde{\sigma}(t) \approx C_\sigma e^{A_\sigma t}B_\sigma$.  For $n
\times n$ $A_\sigma$, we say $(A_\sigma,B_\sigma,C_\sigma)$ is an
\textit{order $n$ realization} of $\widetilde{\sigma}(t)$.  (These
methods may be generalized to non-diagonal $\Gop(t)$ by seeking matrices
$A$, $B$ and $C$ of sizes $n\times n$, $n \times 4$ and $4\times n$
respectively such that $\Gop(t) \approx Ce^{At}B$.)  For
$\widetilde{\sigma}(t) = \sum_i b_i e^{-a_i \gamma t}$, we can exactly
realize $\widetilde{\sigma}(t)$ by choosing $A_\sigma$ diagonal with
entries $-a_i \gamma$, $B_\sigma$ with entries $b_i$, and $C_\sigma =
(1,1,\dots,1)$.  If the $a_i$ are distinct and the $b_i$ non-zero, this
realization is \textit{minimal}: there is no lower-order exact
realization of $\widetilde{\sigma}(t)$.

To find approximate lower-order realizations we use the model reduction
technique of \textit{balanced truncation} \cite{mr_standard, mr_primer}.
Consider a system with time-varying input $u(t) \in \mathbb{R}$, state
$x(t) \in \mathbb{R}^n$, dynamics $\dot{x} = Ax+Bu$, and output $y(t)=Cx
\in \mathbb{R}$; if $u=\delta(t)$, $y=Ce^{At}B$ for $t>0$.  Note that
$(A,B,C) \rightarrow (TAT^{-1},TB,CT^{-1})$ leaves the map
$\Psi:u(t) \mapsto y(t)$ unchanged.  An arbitrary truncation of
state-space dimensions, e.g.  $\left ( \left [
\genfrac{}{}{0pt}{}{a_{11}}{a_{21}}
\genfrac{}{}{0pt}{}{a_{12}}{a_{22}}\right ], \left [
\genfrac{}{}{0pt}{}{b_1}{b_2}\right ], [c_1 \: c_2] \right ) \rightarrow
([a_{11}],[b_1],[c_1])$, may yield a radically different map $\Psi$.
However, we may numerically construct a \textit{balancing}
transformation $T$ such that in the balanced system, a non-negative real
\textit{Hankel Singular Value} (HSV) $h_i$ is associated with each
dimension of the state-space $\mathbb{R}^n$.  Removing all dimensions
with $h_i = 0$ yields a minimal realization; further truncating
dimensions with small HSVs introduces a small error in $\Psi$ which, in
an appropriate norm, is bounded by the sum of the truncated HSVs
\footnote{Balanced truncation for $u = \delta(t)$ will be discussed
elsewhere.}.

\begin{figure}
\includegraphics[scale=0.47]{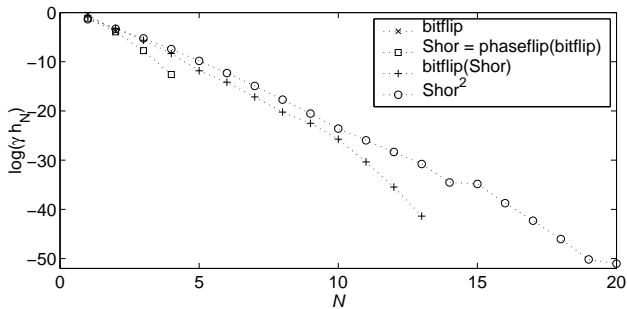}
\caption{Largest HSVs for exact realization of $\widetilde{z}(t)$ at
levels of 3-qubit concatenation (17 smaller values for Shor$^2$ not
shown).}
\label{fig:hsv_exact_Z}
\end{figure}

Writing the series for $\widetilde{\sigma}(t)$ as minimal realizations,
we can balance and calculate their HSVs.  In Fig. \ref{fig:hsv_exact_Z}
we see the HSVs for $\widetilde{z}(t)$ after each level of bitflip and
phaseflip concatenation up to pf(bf(pf(bf))) = Shor$^2$.  Note that
the number of non-zero HSVs grows rapidly at each level of
concatenation, but the number of HSVs above any $h_\mathrm{min}$ grows slowly.
($\widetilde{x}(t)$ and $\widetilde{y}(t)$ give similar results.)

\begin{figure}
\includegraphics[scale=0.47]{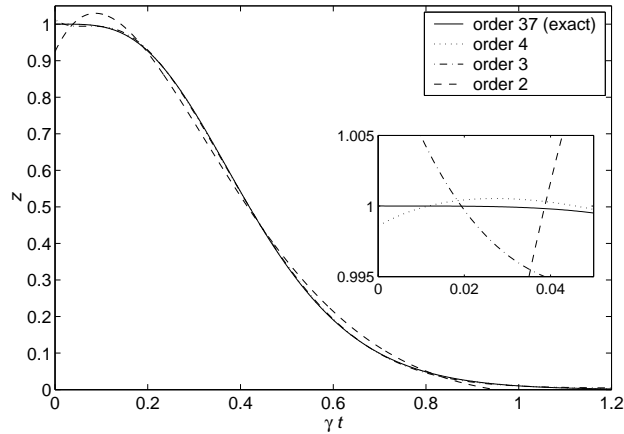}
\caption{Exact $\widetilde{z}_2(t)$, and approximations that result from
balanced truncation.  The order 4 approximation is only distinguishable
from the exact function on the inset.}
\label{fig:baltrunc_Z_inset}
\end{figure}

Consider $\widetilde{z}_2(t)$, with minimal realization of order 37: the
first five HSVs are $(2.5\times 10^{-1})/\gamma$, $(3.7\times
10^{-2})/\gamma$, $(5.3\times 10^{-3})/\gamma$, $(6.0\times
10^{-4})/\gamma$, and $(5.4\times 10^{-5})/\gamma$.  Truncating all but
the four most significant dimensions yields an approximation almost
indistinguishable from the exact $\widetilde{z}_2(t)$; truncating
further to realizations of order 3 and 2 only mildly degrades the
approximation (see Fig. \ref{fig:baltrunc_Z_inset}).

\begin{figure}
\includegraphics[scale=0.47]{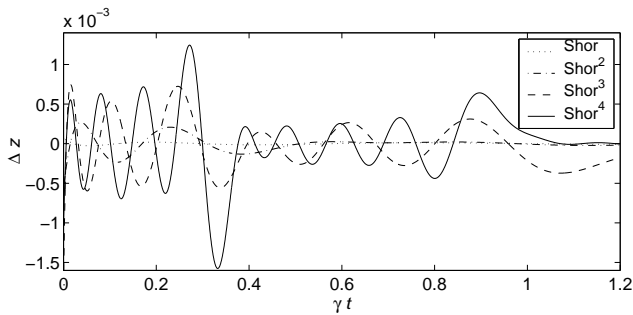}
\caption{Approximation error for $\widetilde{z}_\ell(t)$ generated by
iterative reduction with $h_\mathrm{min} = (4\times
10^{-5})/\gamma$.}
\label{fig:diff_Z}
\end{figure}

Given realizations for the $\widetilde{\sigma}(t)$, we may construct
realizations for polynomials of the $\widetilde{\sigma}(t)$ as follows.
Given $f(t)=C_fe^{A_f t}B_f$ and $g(t)=C_ge^{A_g t}B_g$, the function
$f(t)g(t)$ is realized by $(A_f\otimes\openone + \openone\otimes A_g,
B_f\otimes B_g, C_f \otimes C_g)$.  The function $f(t)+g(t)$ is realized
by $\left ( \left [ \genfrac{}{}{0pt}{}{A_f}{}
\genfrac{}{}{0pt}{}{}{A_g}\right ], \left [
\genfrac{}{}{0pt}{}{B_f}{B_g}\right ], [C_f \:\: C_g] \right )$.  For a
scalar $\alpha$, the function $\alpha f(t)$ is realized by
$(A_f,B_f,\alpha C_f)$.  Composing these operations allows any
polynomial of the $\widetilde{\sigma}(t)$ to be realized, and thus we may
directly apply the $\Omega^C$ to realizations.

For $\ell>2$ it is impractical to construct the exact
$\widetilde{\sigma}_\ell(t)$ and then apply balanced truncation.  Instead, we
build approximate realizations for the $\widetilde{\sigma}_\ell(t)$
using an iterative approach.  Begin with minimal realizations
for the $\widetilde{\sigma}_0(t)$ describing $\Gdep(t)$.  Alternately
apply $\Omega^\mathrm{bf}$ and $\Omega^\mathrm{pf}$ to these realizations;
after each concatenation, balance and truncate dimensions with 
HSVs less than some $h_\mathrm{min}$.  Choosing $h_\mathrm{min} = (4\times
10^{-5})/\gamma$ yields realizations with orders shown in Table
\ref{tbl:terms_and_order}(b).  Comparing to Table
\ref{tbl:terms_and_order}(a), we see the resulting order reduction is
dramatic.

Fig. \ref{fig:diff_Z} shows the differences between the exact
$\widetilde{z}_\ell(t)$ and the results of the iterative reduction
method.  Results for approximating $\widetilde{x}_\ell(t)$ and
$\widetilde{y}_\ell(t)$ are similar.  Up to eight 3-qubit
concatenations, the worst errors $|\Delta\widetilde{\sigma}_\ell(t)|$
are only $\approx 3\times 10^{-3}$.  Note that the errors appear to have
characteristic frequencies; the error is analogous to the ringing in
frequency-limited approximations of step functions.  To good accuracy
the mutual intersection points of the $\widetilde{x}_\ell(t)$ and of the
$\widetilde{z}_\ell(t)$ are preserved; this is expected as the
concatenation polynomials are unchanged.

These results suggest balanced truncation is a powerful approximation
tool in quantum settings.  Future work will further investigate the
iterative reduction method, and attempt to find bounds on the
approximation errors.

\begin{acknowledgments}
This work was partially supported by the Caltech MURI Center for Quantum
Networks and the NSF Institute for Quantum Information.  B.R. acknowledges
the support of an NSF graduate fellowship, and thanks J. Preskill
and P. Parrilo for insightful discussions.
\end{acknowledgments}

\vspace{-4mm}
% Create the reference section using BibTeX:


\begin{thebibliography}{5}
\expandafter\ifx\csname natexlab\endcsname\relax\def\natexlab#1{#1}\fi
\expandafter\ifx\csname bibnamefont\endcsname\relax
  \def\bibnamefont#1{#1}\fi
\expandafter\ifx\csname bibfnamefont\endcsname\relax
  \def\bibfnamefont#1{#1}\fi
\expandafter\ifx\csname citenamefont\endcsname\relax
  \def\citenamefont#1{#1}\fi
\expandafter\ifx\csname url\endcsname\relax
  \def\url#1{\texttt{#1}}\fi
\expandafter\ifx\csname urlprefix\endcsname\relax\def\urlprefix{URL }\fi
\providecommand{\bibinfo}[2]{#2}
\providecommand{\eprint}[2][]{\url{#2}}

\bibitem[{\citenamefont{Nielsen and Chuang}(2000)}]{qecc_standard}
\bibinfo{author}{\bibfnamefont{M.~A.} \bibnamefont{Nielsen}} \bibnamefont{and}
  \bibinfo{author}{\bibfnamefont{I.~L.} \bibnamefont{Chuang}},
  \emph{\bibinfo{title}{Quantum Computation and Quantum Information}}
  (\bibinfo{publisher}{Cambridge University Press}, \bibinfo{year}{2000}),
  \bibinfo{note}{and references therein; J. Preskill, Lecture Notes (1998),
  http://theory.caltech.edu/$\sim$preskill/ph219}.

\bibitem[{\citenamefont{Preskill}(1997)}]{ftc_standard}
\bibinfo{author}{\bibfnamefont{J.}~\bibnamefont{Preskill}}
  (\bibinfo{year}{1997}), \eprint{quant-ph/9712048}.

\bibitem[{\citenamefont{Dullerud and Paganini}(2000)}]{mr_standard}
\bibinfo{author}{\bibfnamefont{G.~E.} \bibnamefont{Dullerud}} \bibnamefont{and}
  \bibinfo{author}{\bibfnamefont{F.~G.} \bibnamefont{Paganini}},
  \emph{\bibinfo{title}{A Course in Robust Control Theory}}
  (\bibinfo{publisher}{Springer-Verlag}, \bibinfo{year}{2000}).

\bibitem[{\citenamefont{Rahn}()}]{mr_primer}
\bibinfo{author}{\bibfnamefont{B.}~\bibnamefont{Rahn}}
  (\bibinfo{year}{2001}), \eprint{quant-ph/0112066}.

\bibitem[{\citenamefont{Devaney}(1989)}]{dynamical_standard}
\bibinfo{author}{\bibfnamefont{R.~L.} \bibnamefont{Devaney}},
  \emph{\bibinfo{title}{An Introduction to Chaotic Dynamical Systems}}
  (\bibinfo{publisher}{Addison-Wesley}, \bibinfo{year}{1989}).

\end{thebibliography}
\end{document}